# The New Superconductor tP-SrPd$_2$Bi$_2$: Structural Polymorphism and Superconductivity in Intermetallics

Weiwei Xie*, Elizabeth M. Seibel, and Robert J. Cava*

Department of Chemistry, Princeton University, Princeton, NJ, USA, 08540

*Supporting Information*

**ABSTRACT:** We consider a system where structural polymorphism suggests the possible existence of superconductivity through the implied structural instability. SrPd$_2$Bi$_2$ has two polymorphs which can be controlled by the synthesis temperature: a tetragonal form (CaBe$_2$Ge$_2$-type) and a monoclinic form (BaAu$_2$Sb$_2$-type). Though the crystallographic difference between the two forms may at first seem trivial, we show that tetragonal SrPd$_2$Bi$_2$ is superconducting at 2.0 K whereas monoclinic SrPd$_2$Bi$_2$ is not. We rationalize this finding and place it in context with other 1-2-2 phases.

Superconductivity is difficult to predict successfully from first principles in new materials, and thus our recent work has concentrated on developing empirical chemical guidelines to search for new superconductors.[1,2,3] The relationship between structural instability and conventional superconductivity, manifested through structural phase transitions at low temperatures, has long been discussed from the physics perspective in intermetallic compounds such as the A15 and Laves phases.[4] Similarly, superconductivity has recently been found in many iron pnictides in the shadow of a suppressed low temperature structural phase transition in their ThCr$_2$Si$_2$ structure type hosts; even though the mechanism for the superconductivity appears to involve much more than electron-lattice coupling, it is not likely that the superconductivity and structural phase transitions are unrelated.[5] Thus knowing that superconductivity in the ThCr$_2$Si$_2$ structure type can occur near a structural transition, we set out to explore further the concept that superconductivity and structural instability can go hand-in-hand in intermetallics by considering this relationship from a different perspective: for a case where structural instability is reflected through polymorphism rather than through the presence of a temperature-dependent structural phase transition. Here we report the realization of this design principle through a study of polymorphic SrPd$_2$Bi$_2$: we find that tetragonal primitive (tP) SrPd$_2$Bi$_2$ is superconducting above 2 K and monoclinic primitive (mP) SrPd$_2$Bi$_2$ is not.

SrPd$_2$Bi$_2$ has been reported as polymorphic, existing in both the tetragonal primitive CaBe$_2$Ge$_2$ and monoclinic primitive BaAu$_2$Sb$_2$ structure types, depending on differences in the synthesis temperature (Figure 1).[6] The tetragonal primitive form of SrPd$_2$Bi$_2$, tP-SrPd$_2$Bi$_2$, is closely related to the ThCr$_2$Si$_2$ structure type of the superconducting iron pnictides. Despite their crystallographic relationship, superconductors in the ThCr$_2$Si$_2$ structure type are relatively common but those in the CaBe$_2$Ge$_2$-type structure are rare.[7] While the ThCr$_2$Si$_2$ structure has two equivalent M$_2$X$_2$ layers per cell, identical by symmetry through the body centering, the CaBe$_2$Ge$_2$ structure consists of alternating M$_2$X$_2$ and X$_2$M$_2$ layers, which are independent in the primitive tetragonal cell. This structure is also related to that of LaFeAsO, another host for pnictide superconductivity (Figure 1).[7]

The origin of the polymorphism in SrPd$_2$Bi$_2$ is a symmetry-lowering twist of the constituent tetrahedra (Figure 1), which we argue impacts the existence of superconductivity in a manner similar to the temperature-induced structural phase transition that occurs in the superconducting ThCr$_2$Si$_2$-type iron arsenides. The iron arsenide superconductors are often discussed from the perspective of ionic M-X charge transfer; the same is not the case, however for compounds in the CaBe$_2$Ge$_2$-type structure. In the CaBe$_2$Ge$_2$-type structure M and X are typically heavy elements with a small electronegativity difference, so the distinction between transition metals and main group elements is not so clear. In the current case of SrPd$_2$Bi$_2$, there is no clear electronegativity differentiation between Pd and Bi: According to Pauling's scale, the metal Pd (2.20) is more electronegative than Bi (2.02), while for Mulliken's or Pearson's scales the opposite is true, with Bi (2.02) more electronegative than Pd (1.59).[8,9] As a consequence, SrPd$_2$Bi$_2$ should be considered an intermetallic compound, with a more subtle form of internal charge redistribution than is found in ThCr$_2$Si$_2$-type compounds. In other words, it is not one where the type of charge transfer involved is predominately ionic such as is found in oxides or classical Zintl phases.[10]

Polycrystalline pellets of SrPd$_2$Bi$_2$ were synthesized via solid state reaction. Stoichiometric amounts of Pd (powder, 99.99%, Alfa Aesar), and Bi (powder, 99.999%, Alfa Aesar) were mixed with 50% molar excess Sr (in order to balance Sr loss during synthesis; pieces, 99.9%, Alfa Aesar), and compacted into 300 mg pellets in a hydraulic press. The pellets were placed inside alumina crucibles and heated in evacuated quartz ampoules at 1050°C for at least 20 h, then reground, re-pelletized, and re-heated for at least two more nights. The SrPd$_2$Bi$_2$ samples were examined by powder X-ray diffraction for identification and phase purity on a Bruker D8 ECO powder diffractometer employing Cu K$\alpha$ radiation. Quantitative phase analysis was checked through a full-profile Rietveld refinement with Rietica. The quantitative analysis of the powder diffraction pattern (FIGURE 2) shows that the pure polycrystalline sample employed for the bulk property characterization matches the previously reported tetragonal primitive SrPd2Bi2.

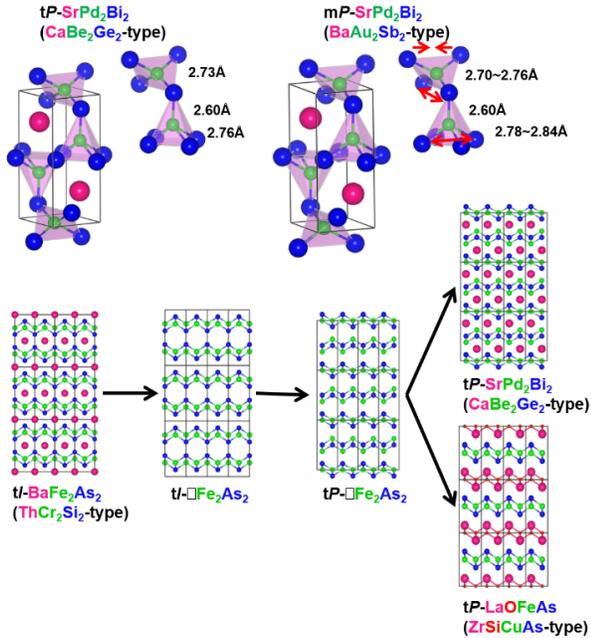

**FIGURE 1. The Polymorphism in SrPd2Bi2.** (Upper panel) the relationship between the CaBe2Ge2 and BaAu2Sb2-type structures for SrPd2Bi2 (Lower panel) Schematic diagram of the relationship between the body-centered tetragonal ThCr2Si2-type structure of BaFe2As2 and the primitive tetragonal structures of SrPd2Bi2 (CaBe2Ge2 type) and LaOFeAs (or LaFeAsO - ZrCuAsSi type).

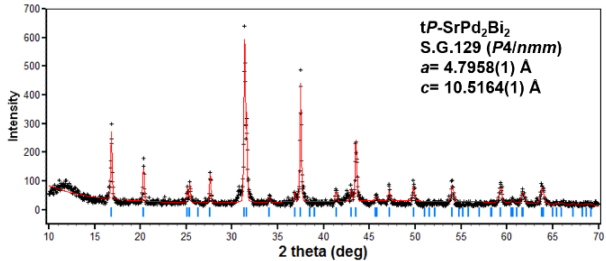

**FIGURE 2. Powder X-ray diffraction pattern of SrPd2Bi2 showing the pure CaBe2Ge2-type SrPd2Bi2 compound studied.** The red solid line shows the corresponding Rietveld fitting. Blue tick marks are the expected peak positions.

The temperature (T) dependent electrical resistivity of tP-SrPd2Bi2 between 0.4 and 300 K is shown in Figure 3(a). After displaying typical 'poor metal' behavior on cooling (i.e. where the resistivity does not drop much with decreasing temperature), the resistivity undergoes a sudden drop to zero at 2.2 K, indicating the presence of superconductivity. In correspondence with ρ(T), the magnetic susceptibility (χ(T)) shown in Figure 3 (b) begins its decrease suddenly on cooling at ~2.0 K and shows large negative values. The transition is not complete by the minimum temperature of the measurement apparatus (1.8 K), but already shows a volume susceptibility of 60% of the full possible Meissner volume fraction, indicating that the diagmagnetic signal is too large for the observed superconductivity to arise from small amounts of spurious phases.[11] We also measured mP-SrPd2Bi2 and find no intrinsic diamagnetism characteristic of superconductivity.

To further prove that the observed superconductivity is intrinsic to only tP-SrPd2Bi2, the superconducting transition was characterized through specific heat measurements.[12] These measurements are a reliable indication of the presence of bulk superconductivity when combined with resistivity and susceptibility due to their characterization of the change in bulk thermodynamic properties at the superconducting transition. The specific heat for zero-field and field-cooled tP-SrPd2Bi2 in the temperature range of 1.85 K to 10K is presented in Figure 3(b)(insert). The good quality of the sample and the bulk nature of the superconductivity are strongly supported by the presence of a large anomaly in the specific heat at Tc= 1.9 ~ 2.1 K, in excellent agreement with the Tc determined by ρ(T) and χ(T). The electronic contribution to the specific heat (γ) was measured in a field of 5 T to suppress the superconductivity and found to be 6.12 mJ/mol-K2. (The data are fitted using the formula $C_p = \gamma T + \beta T^3$, in which γ and β are the electronic and lattice contributions to the specific heat, respectively.) The value of the specific heat jump at Tc, estimated by the maximum value of specific heat observed at the low temperature limit of the measurement, is consistent with that expected from a weak-coupling BCS superconductor; $\Delta C_{el}/\gamma T_c$ per mole SrPd2Bi2 in the pure sample = 1.27.[13] This ratio is already within error of the BCS superconductivity weak coupling value of 1.43, and is in the range observed for many superconductors. As an added check, we tested PdBi down to 1.85 K and found that it is not superconducting and therefore could not give rise to the observed specific heat feature. Thus the observed superconductivity is intrinsic to tP-SrPd2Bi2.

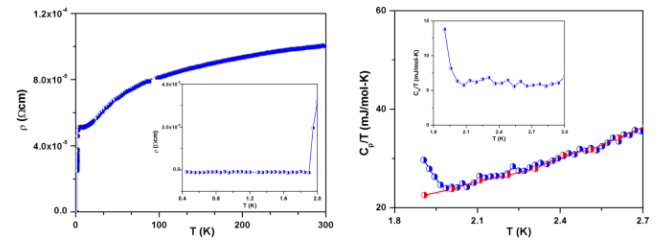

**FIGURE 3. Property characterization of the tP-SrPd2Bi2 superconductor.** (a) The temperature dependence of the electrical resistivity of tP-SrPd2Bi2 showing the superconducting transition near 2K. (Insert) The resistivity ranging from 10K to 1K. (b) (Main panel) The temperature dependences of the magnetic susceptibility for tP-SrPd2Bi2 and mP-SrPd2Bi2 between 1.8K and 3K in an applied field of 10 Oe after zero-field cooling. (Insert) Specific heat characterization of the superconducting transition of tP-SrPd2Bi2. Temperature dependence of the specific heat Cp of a tP-SrPd2Bi2 sample measured with (μoH = 5T) and without a magnetic field.

To gain further insight into the relationship between the polymorphism and the electronic states of SrPd2Bi2, we inves-tigated the electronic density of states (DOS) and band structures for both tP- and mP-SrPd2Bi2 calculated from first principles using WIEN2k.[14,15] We first focus on the DOS comparison between the two phases. The calculated DOS curves are illustrated in Figure 4 (upper panel), which emphasizes contributions from -2eV to +2eV of $E_F$. The analysis of the orbitals contributing to the bands from band structures at $E_F$ shows the dominance of the 4d states in Pd and 6s and 6p states in Bi. The significant DOS at $E_F$ in both cases is consistent with the metallic properties. A peak in the DOS like the one in the DOS near $E_F$ for tP-SrPd2Bi2 is often taken to be an indication of a near-by structural, electronic, or magnetic instability such as superconductivity. In contrast, there is no such clear, isolated DOS peak for the monoclinic polymorph. The peak in the DOS for tP-SrPd2Bi2 is due to the presence of saddle points in the electronic structure at the Z and X points in the Brillouin zone (Figure 4 lower). These saddle points near $E_F$ are often proposed to be important for yielding superconductivity in a variety of materials. The symmetry-lowering in mP-SrPd2Bi2 from the twist of the MX4 tetrahedra causes the saddle points that occur in tP-SrPd2Bi2 to vanish, and the bands split and cross the Fermi level. This demonstrates that polymorphism, which at first glance may appear to be reflected in only a small crystallographic difference, can have critical implications on the electronic structure in the same way that temperature-induced structural phase transitions do.

In conclusion, the tetragonal CaBe2Ge2-type and monoclinic BaAu2Sb2-type polymorphism in SrPd2Ge2 provide a good system for examining the hypothesis that structural polymorphism may be a significant indicator for possible superconductivity in intermetallic compounds. Resistivity and heat capacity measurements show only tP-SrPd2Bi2 to be a superconductor with a Tc ~2.0 K. The work described here suggests that considering compounds that display structural polymorphism may be a successful design paradigm for discovering new intermetallic superconductors.

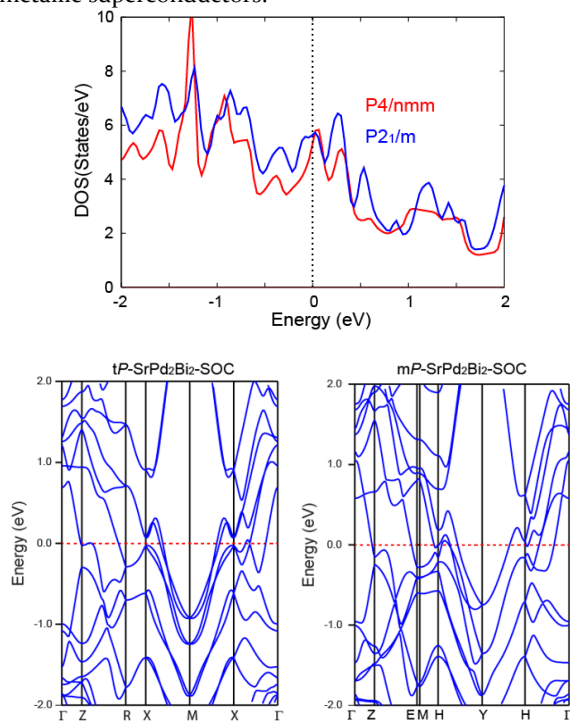

FIGURE 4. Results of the Electronic structure calculations with emphasis on the vicinity of the Fermi Energy. (Left) CaBe2Ge2-type and (right) BaAu2Sb2-type SrPd2Bi2. Total DOS curves and band structure curves are obtained from non-spin-polarized LDA calculations with spin orbit coupling. (Up) Density of states. (Bottom) Band structures of two phases

AUTHOR INFORMATION

Corresponding Author

W. Xie (weiweix@princeton.edu)
R.J.Cava (rcava@princeton.edu)

Notes
The authors declare no competing financial interests.

ACKNOWLEDGMENT

The resistivity measurement and electronic structure calculations were supported by the Department of Energy, grant DE-FG02-98ER45706. The powder X-ray powder diffraction data acquisition and analysis, and the specific heat measurements and analysis, were supported by the Gordon and Betty Moore Foundation's EPiQS Initiative through Grant GBMF4412.

Table of Content

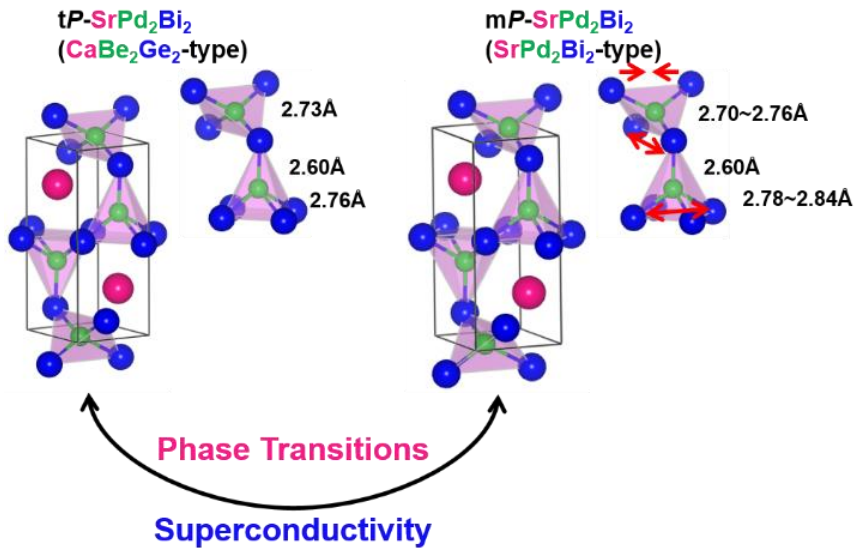

Structural polymorphism suggests the possible existence of superconductivity through the implied structural instability. $SrPd_2Bi_2$ has two polymorphs which can be controlled by the synthesis temperature: a tetragonal form ($CaBe_2Ge_2$-type) and a monoclinic form ($BaAu_2Sb_2$-type. We show that tetragonal $SrPd_2Bi_2$ is superconducting at 2.0 K whereas monoclinic $SrPd_2Bi_2$ is not.